\documentclass[twocolumn,superscriptaddress,showpacs,amsmath,amssymb,nofootinbib]{revtex4-1}

\usepackage{graphicx}
\usepackage{dcolumn}
\usepackage{bm}
\usepackage{float}

\begin{document}

\title{Single-shot MeV transmission electron microscopy with picosecond temporal resolution }

\author{R. K. Li}
\author{P. Musumeci}
\affiliation{Department of Physics and Astronomy, UCLA, Los Angeles,
CA, 90095, USA}

\date{\today}

\begin{abstract}
Pushing the limits in temporal resolution for transmission electron microscopy (TEM) requires a revolutionary change in the electron source technology. In this paper we study the possibility of employing a radiofrequency photoinjector as the electron source for a time-resolved TEM. By raising the beam energy to the relativistic regime we minimize the space charge effects which otherwise limit the spatio-temporal resolution of the instrument. Analysis and optimization of the system taking into account the achievable beam brightness, electron flux on the sample, chromatic and spherical aberration of the electron optic system, and space charge effects in image formation are presented and supported by detailed numerical modeling. The results demonstrate the feasibility of 10 nanometer - 10 picosecond spatio-temporal resolution single-shot MeV TEM.
\end{abstract}


\maketitle

\section{Introduction}

Transmission electron microscopy (TEM) is an extremely powerful and versatile tool in many research areas, including biological, chemical and material sciences, as well as for industrial applications~\cite{tem1,tem2,tem3}. For decades one of the main efforts in improving TEM has been perfecting the spatial resolution. High voltage ($>1$ MeV) electron microscopy had originally been developed to take advantage of the shorter wavelength of higher energy electrons to beat the limits set by lens aberrations and approach sub-atomic resolution~\cite{hvtem}. Very recently spherical and chromatic aberration correction modules were successfully developed and sub-{\AA}ngstrom spatial resolution has been achieved at 100-300 kV electron beam energies~\cite{Haider:Nature1998, Krivanek:Nature1998}.

An ongoing trend in further extending the scope of the research that can be performed with TEM is to add the temporal dimension to the instrument capabilities thus enabling the real-time observation of microscopic dynamical processes~\cite{king05,futureneeds}. One of the first successful attempts to develop a high-speed TEM at TU-Berlin produced an instrument capable of taking 10 nanosecond (ns) temporal snapshots of the samples with a few hundred nanometers (nm) spatial resolution ~\cite{Bostanjoglo02,Bostanjoglo:rsi03}. More recently, the dynamic transmission electron microscopy (DTEM) at LLNL~\cite{dtem,dtem:science08} demonstrated single-shot images of samples with 10 ns temporal resolution and 10 nm spatial resolution. At the relatively low energy (200 keV) of this instrument electron-electron ($e$-$e$) interactions in the lenses cross-overs prevent the use of a beam current higher than a few mA and limit the spatio-temporal resolution. The ultrafast electron microscopy (UEM) technique developed at Caltech employs a different scheme - the stroboscopic method - using on average one electron per pulse to record images of reversible dynamic processes with $\sim$500 fs temporal resolution ~\cite{caltechuem,zewailbook}. Because single-electron packets have no space-charge broadening, the problem is eliminated at its roots and the spatial resolution can approach a level similar to that achieved in conventional TEMs. On the other hand, millions or more electron pulses are needed to take each image with the repetition rate typically limited to the MHz range to allow enough time for the sample to relax to its initial state. With this technique one is restricted to dynamical processes occurring in the same exact way for each pump-probe cycle.

In order to overcome these limitations and push the boundaries in spatio-temporal resolution in electron-based imaging, a radically new approach to the electron source for the TEM is needed. In this paper we discuss the application of radiofrequency (RF) photoinjectors to single-shot picosecond temporal resolution MeV TEM. Among various types of high brightness electron beam sources, photocathode RF guns have the unique capability of very high acceleration gradient ($\sim$100 MV/m), high final beam energy ($\sim$3-5 MeV), and high beam charge (up to a few nC or 10$^{10}$ particles per pulse)~\cite{dowell:book}. These sources have enabled the revolutionary advent of X-ray free-electron lasers~\cite{LCLS:commissioning}, but have not been employed yet in electron-based imaging systems. Nevertheless their potential for electron-scattering techniques has been widely recognized and the last decade has witnessed the exciting progress of using photocathode RF guns for ultrafast electron diffraction \cite{xjued,hastings,Musumeci:APL,lrkrsi09,osakaued,bnlued13,regae} and more recently for initial exploration of time-resolved TEM~\cite{yanguem}.

In this paper we will consider various components of the system and present a complete feasibility study for a novel time-resolved MeV TEM capable of taking single-shot images with 10 nm - 10 ps spatio-temporal resolution, improving by three orders of magnitude on current state-of-the-art performances. In section II, we will analyze the requirements on beam quality showing the motivations behind the use of MeV beams for this application, independently on the particular design of the electron source and electron optics. In Section III we present an optimization of the RF photoinjector beam to achieve transverse brightness and energy spread well beyond the state-of-the-art which fulfills the sample illumination requirements. In Section IV, we discuss the optical design of the column downstream of the sample, and evaluate a novel scheme based on the use of quadrupole magnets as objective and projector lenses. This solution offers a convenient alternative to solenoidal round lenses, which quickly become large and expensive when the beam energy is scaled up to relativistic levels. In particular we evaluate the aberrations and discuss the optimization of a permanent-magnet-based quadrupole triplet with effective focal length $<2$ cm. Finally, in Section V, we consider the effects of $e$-$e$ interactions including both smooth space charge forces and stochastic scattering. With the help of self-consistent numerical modeling we simulate the image formation process in the column for an idealized test sample. The results confirm the feasibility of capturing 10 nm resolution single-shot images using 10 ps long electron beams.

\section{General considerations for single-shot UEM}

We begin our discussion with the estimate for the required electron flux to distinguish sample features having 50$\%$ contrast. Applying the Rose criterion \cite{rose} to evaluate the minimum number of particles to generate a clear signal above the noise induced by the Poisson statistics, we find the need for 100 or more electrons per spatial resolution unit $d$ to maintain the shot-noise below 10\% and achieve a signal-to-noise ratio of 5. Therefore, rougly $4\times10^6$ electrons (0.6 pC) are required to form an image with $4\times10^4$ resolution units with an area of $200d\times200d$. If we want to pack this amount of electrons in a single pulse with 10 ps bunch length, the peak current in the microscope column will be 60 mA or larger.

For these large peak currents, several orders of magnitude higher than those in conventional TEMs and 10-100 times larger than those in DTEM, there are a number of reasons to increase the electron energy to the MeV level. Firstly, the higher electron energy significantly eases the issue related to stochastic Coulomb scattering. This effect can not be compensated for by simply increasing the strength of the lenses, since random collisions essentially heat up the phase space. The effect of charged particle collisions in a crossover has been studied analytically by Jansen~\cite{jansen,jansenbook}. More recently Reed and collaborators~\cite{bryan09} have analyzed numerically its consequences for time resolved electron microscopy. By performing particle simulations for different beam energies with a pairwise model for $e$-$e$ interactions, they found the blur on the final image to be negligible only provided the beam energy is above a few MeV.

Secondly, relativistic electron sources are typically characterized by very high extraction fields at the cathode, compared to conventional TEMs which are limited by arcing in the gun to gradients smaller than 10 MV/m \cite{gunreview}. The higher electric fields at the cathode allow (for a given beam charge) to decrease the source size and improve beam brightness. Recent analysis shows that for an initial cigar aspect-ratio (long and narrow pulse) a relatively large charge can be extracted from a small transverse region of the cathode. For example, more than 1 pC charge could be extracted from 10 $\mu$m spot size with a bunch length of 10 ps in the 100 MV/m peak field of an $S$-band photocathode RF gun~\cite{filippetto}.

Besides the challenging demand on the beam flux, small transverse emittance as well as very low energy spread are also critical to minimize the effects of spherical and chromatic aberrations and reach 10 nm spatial resolution. The required 6D phase space density greatly depends on the type of imaging mode that is planned for the microscope. For incoherent imaging the resolution can be estimated using
\begin{equation}
d = \sqrt{\bigl(\lambda / \beta \bigr)^2 + (C_s \beta^3)^2 + (C_c \beta \delta\gamma/\gamma)^2}
\label{resolution}
\end{equation}
where $\beta$ is the objective aperture collection semi-angle and $\delta\gamma/\gamma$ is the relative beam energy spread.

For simplicity, we can assume that the resolution is far above the diffraction limit (reasonable since for MeV electrons the de Broglie wavelength $\lambda<1$ pm) and neglect the first term under the square root. It is then straightforward to obtain simple estimates for acceptable beam parameters from the spherical and chromatic aberration coefficients ($C_s$ and $C_c$ respectively). These are related to the $U_{1222}$ and $T_{126}$ elements of the third and second order beam transfer matrices. As a first approximation these have values comparable with the lens focal distance which for MeV electrons can be as short as 2 cm. It then follows from Eq.~\ref{resolution} in order to get 10 nm spatial resolution we will need an relative energy spread lower than $10^{-4}$ and a collection semi-angle $\lesssim 5$ mrad. Actually, in order to increase the contrast it is preferable to have a smaller divergence, at most comparable with the Bragg angle which is usually 1 mrad for MeV electrons.

Combined with the Rose criterion we can use the beam divergence (equal to the collection angle in bright-field imaging mode) as an upper bound to estimate the required transverse emittance. With a density of $100e$/(10 nm)$^2$ and a total charge of 0.6 pC the rms spot size of the full beam at the sample must be smaller than $0.5~\mu$m. Note that in this case the dose on the sample will be lower than the damage threshold \cite{Egerton:micron}. As we push the spatial resolution to 1 nm and beyond, dealing with the damage will become the limiting factor in the microscope design. In principle if sub-ps temporal resolution were to become feasible, one could attempt to use an ultrafast probe to outrun the damage following the diffract-and-destroy scheme successfully demonstrated in XFELs~\cite{outrun}.

For a $\gamma = 10$ beam with an rms beam divergence $\sigma_\theta=1$ mrad the rms normalized emittance is $\epsilon_n=5$ nm. With these parameters, the coherence length of the beam $L_c = \lambda/2\pi \sigma_\theta$ is $<1$ {\AA} so no coherent imaging will be possible. The initial operation mode of our single-shot ps MeV TEM is limited to incoherent mass-thickness contrast imaging. Other contrast mechanisms might be enabled by future improvements on the beam emittance using ultralow thermal emittance cathodes~\cite{dowell3step,Dowell:photocathodes}. A substantial progress on the beam quality (both in terms of emittance and energy spread) will be required in order to enable single-shot high temporal resolution coherent imaging.

\begin{table}[h]
\caption{Requirements on electron source parameters.}
\begin{ruledtabular}
\begin{tabular}{ccc}
& RF photogun & ps MeV TEM \\
Number of electrons & 10$^7$ & $>10^6$ \\
rms normalized emittance & 40 nm & $<10$ nm \\
rms energy spread & 10$^{-3}$ & $<10^{-4}$ \\
FWHM bunch length & $<200$ fs & 10 ps \\
\end{tabular}
\end{ruledtabular}
\label{table1}
\end{table}

In Table~\ref{table1}, we report the best performances of state-of-the-art RF guns \cite{nanoemittance,beamreview} and the requirements for a 10 nm - 10 ps imaging system validating the use of this source for time-resolved electron microscopy. One of the main challenge lies in the control of the energy spread. In the optimization discussed in the next section we will trade off bunch length to improve transverse emittance and energy spread enabling 10 nm spatial resolution imaging.

\section{Generation of the electron beams}

Guided by the general considerations presented in the previous section, here we discuss the optimization of an RF photoinjector design aimed at generating electron beams suitable for single-shot ps TEM.

\begin{figure}[htb]
\includegraphics*[width=85 mm]{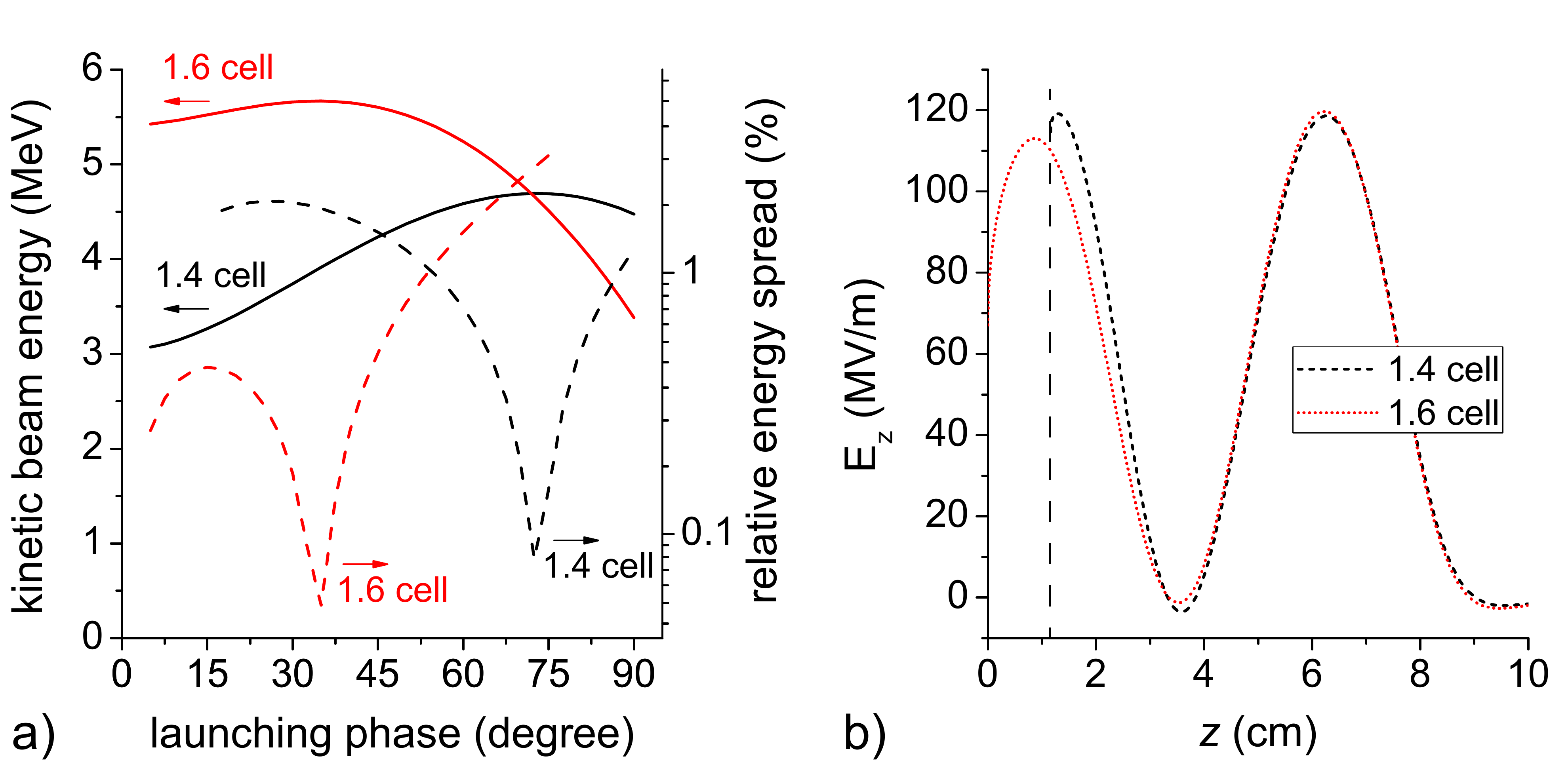}
\caption{(Color online) Comparison of 1.6 cell and 1.4 cell guns for the (a) output kinetic energy and energy spread (10 ps long, low charge electron beam) and (b) the acceleration gradient seen by the beam.}
\label{1p4cellgun}
\end{figure}

Due to the limits in the charge emission set by the initial charge density, the transverse beam brightness $B_{4D}$ scales at least linearly with the extraction field $E_0$, the accelerating field at the cathode during emission~\cite{ivan09,filippetto}. For a pancake beam aspect-ratio we have $B_{4D} \propto E_0$ while for cigar beams the dependence is even stronger. In both cases, though, it is of panamount importance to maximize $E_0$. For this reason, we propose a novel 1.4 cell $S$-band photocathode RF gun structure instead of a more commonly used 1.6 cell type. The longer half cell length was used in early designs to maximize the output energy and minimize the defocusing kick at the exit iris. This helped to control the emittance for 1 nC mm-size beams, but it is not required for the much smaller beam sizes and charges needed for the microscopy application. On the contrary, by shortening the length of the first cell the optimal injection phase (for maximum output energy and minimum energy spread) shifts from the typical 25$^\circ$-35$^\circ$ range to around 70$^\circ$-75$^\circ$ (see Figure~\ref{1p4cellgun}). The extraction field is thus increased by a factor of $\sin(70^\circ)/\sin(30^\circ)=1.9$. For gun operating at $120$ MV/m gradient, the extraction field will then be $E_0>110$ MV/m. The larger extraction field eases the space charge effects at emission and in the propagation region close to the cathode, allowing very high current densities from a small area, minimizing the space charge induced emittance growth~\cite{ivan09}.

The small photoemission source area (a few tens of $\mu$m radius) is illuminated in our design by a long ultraviolet (UV) laser pulse to generate a cigar-aspect ratio electron beam. The initial transverse beam distribution is tailored to a spherical profile by properly shaping the UV laser pulse. If the initial longitudinal profile of the electron beam is parabolic, the beam will expand transversely driven by strong space-charge forces to form a nearly ideal uniformly filled ellipsoid. The final beam distribution is characterized by very linear phase spaces~\cite{Claessens05,nanoemittance} and the beam emittance can be preserved close to its initial value~\cite{luiten04,pietro08,zhou12}. As we will discuss in Section V uniform charge density at the sample is required to maintain uniform space charge defocusing along the beam. This can be achieved by employing a flat-top longitudinal profile. Since for very large aspect-ratios the transverse and longitudinal dynamics are essentially uncoupled, space charge forces cause a strong transverse expansion leaving the longitudinal profile essentially frozen. For the simulations in this paper we used a 10 ps full width flat-top laser profile with fast (100 fs) rising and falling edges. An initial emittance of 10 nm-rad was obtained by assuming an intrinsic emittance of 0.5 mm-mrad per mm rms initial spot size. This value for intrinsic emittance has been experimentally demonstrated \cite{hauri10} and can be further improved by lowering the UV photon energy to less than 0.1 eV above the effective cathode work function.

\begin{figure}[htb]
\includegraphics*[width=85 mm]{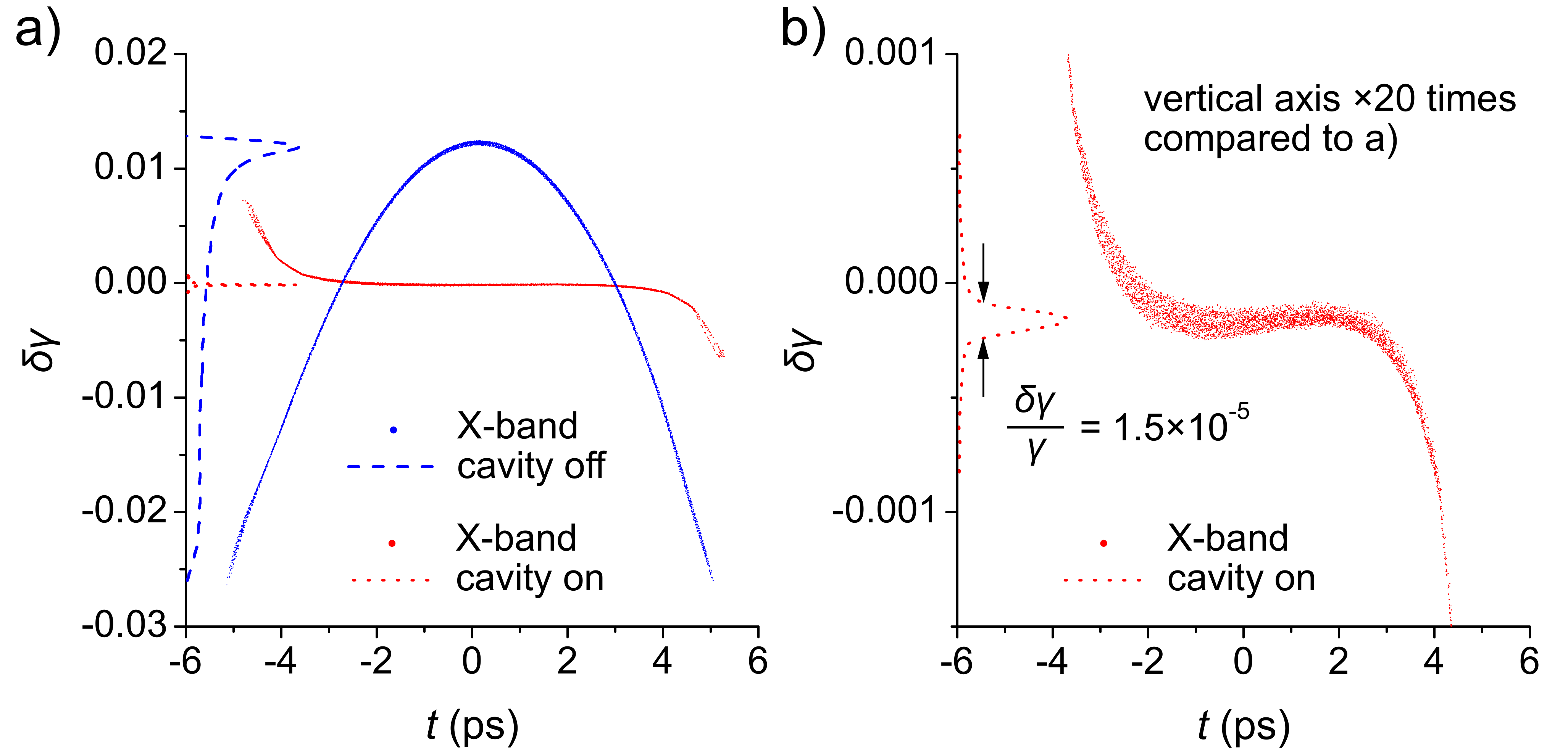}
\caption{(Color online) (a) Longitudinal phase space (LPS) of the beam before (blue dot) and after (red dot) the $X$-band cavity. The correlated energy spread $\delta\gamma/\gamma$ is reduced by 150 times. (b) LPS at the sample position. The vertical axis is zoomed-in by 20 times compared to that of (a).}
\label{figlps}
\end{figure}

An important characteristic of such a long beam from an $S$-band RF gun is a relatively large energy spread. The longitudinal phase space (LPS) distribution at the gun exit has a strong RF curvature induced correlation as shown in Fig.~\ref{figlps}(a). The total beam energy spread is much larger than the uncorrelated width of the LPS distribution.

For a uniform 10 ps long bunch (extending for $\Delta \phi$ = 10$^\circ$ of $S$-band phase) we can calculate the rms energy spread at gun exit from the expression for the output beam energy
\begin{equation}
\gamma_f=\gamma_0 \cos(\phi)
\end{equation}
where $\phi = 2 \pi f_0 t$ is the output phase of the particles, $\gamma_0$ is the maximum energy, and $f_0=2.856$ GHz is the gun frequency~\cite{xj96}. We have $\gamma_0=10.19$ and $\delta \gamma_f / \gamma_f$ = 10$^{-3}$, in good agreement with the General Particle Tracer (GPT)~\cite{gpt} simulation results in which space charge effects are included, as shown in Fig.~\ref{figlps}, and much larger than what required for imaging application. This quantity is proportional to $\Delta \phi^2$, thus in order to reduce it to an acceptable level one possibility is to use a $<$1 ps long beam at the cathode. But in this case, the peak current would then be 10 times larger resulting in unbearable space-charge effects and emittance degradation.

We analyze a different solution involving the use of an additional RF cavity as an energy spread compensator. At first order the LPS distribution is characterized by an almost ideally quadratic curvature which can be compensated using a higher harmonic RF cavity operating at the opposite deceleration phase. As an example we consider $f_1=4f_0=11.424$ GHz $X$-band cavity and the final beam energy after the compensator cavity is
\begin{equation}
\gamma_c=\gamma_0 \cos(2 \pi f_0 t) - \gamma_1 \cos (2 \pi f_1 t)
\end{equation}
where $\gamma_1$ is the maximum energy gain of the $X$-band cavity. To cancel the quadratic term we need $\gamma_1=\gamma_0/(f_1/f_0)^2 \approx0.6$. Such a small amount of energy loss can be realized by using a single-cell (length $L_X=0.83$ cm) $X$-band cavity at a few tens of MV/m gradient. The compensator cavity will be placed immediately following the $S$-band gun and before the condenser lens to minimize chromatic effects. Previously proposed two-frequency RF gun is a similar concept and might fulfill the beam requirement for single-shot ps TEM~\cite{twofregun1,twofregun2,twofregun3,twofregun4}.

In Fig.~\ref{figlps}(a) and (b) we show the compensated LPS (red dots), and the correlated beam energy spread $\delta\gamma/\gamma$ is reduced to low $10^{-5}$ level (FW50 definition, full width containing 50\% of the particles), in agreement with an analytical estimate of the residual energy spread from the higher order terms in the RF curvature (proportional to $\Delta \phi^4$). Most of the final beam energy spread is due to residual longitudinal as well as transverse space charge forces. We note that the success of the RF curvature compensation scheme relies on precise and high stability control of the amplitudes and phases of the $S$-band and $X$-band RF sources, enabled by recent advances on RF-laser synchronization~\cite{laserrftiming1,laserrftiming2} and high stability modulator technologies~\cite{modulator1,modulator2}.

Once a beam with the correct phase space properties (energy spread and transverse emittance) is generated, the next component in the microscope is the transport optics required to illuminate the sample under optimal conditions. This function is typically performed by one or more condenser stages in a TEM. In this paper we restrict the discussion to only a single-lens condenser stage instead of a more flexible multi-lenses design~\cite{bryan10} in order to minimize the number of beam waists (high charge density regions) in the system. A two-lens condenser stage could be considered for improving the flexibility of operation.

\begin{figure*}[htb]
\includegraphics*[width=140 mm]{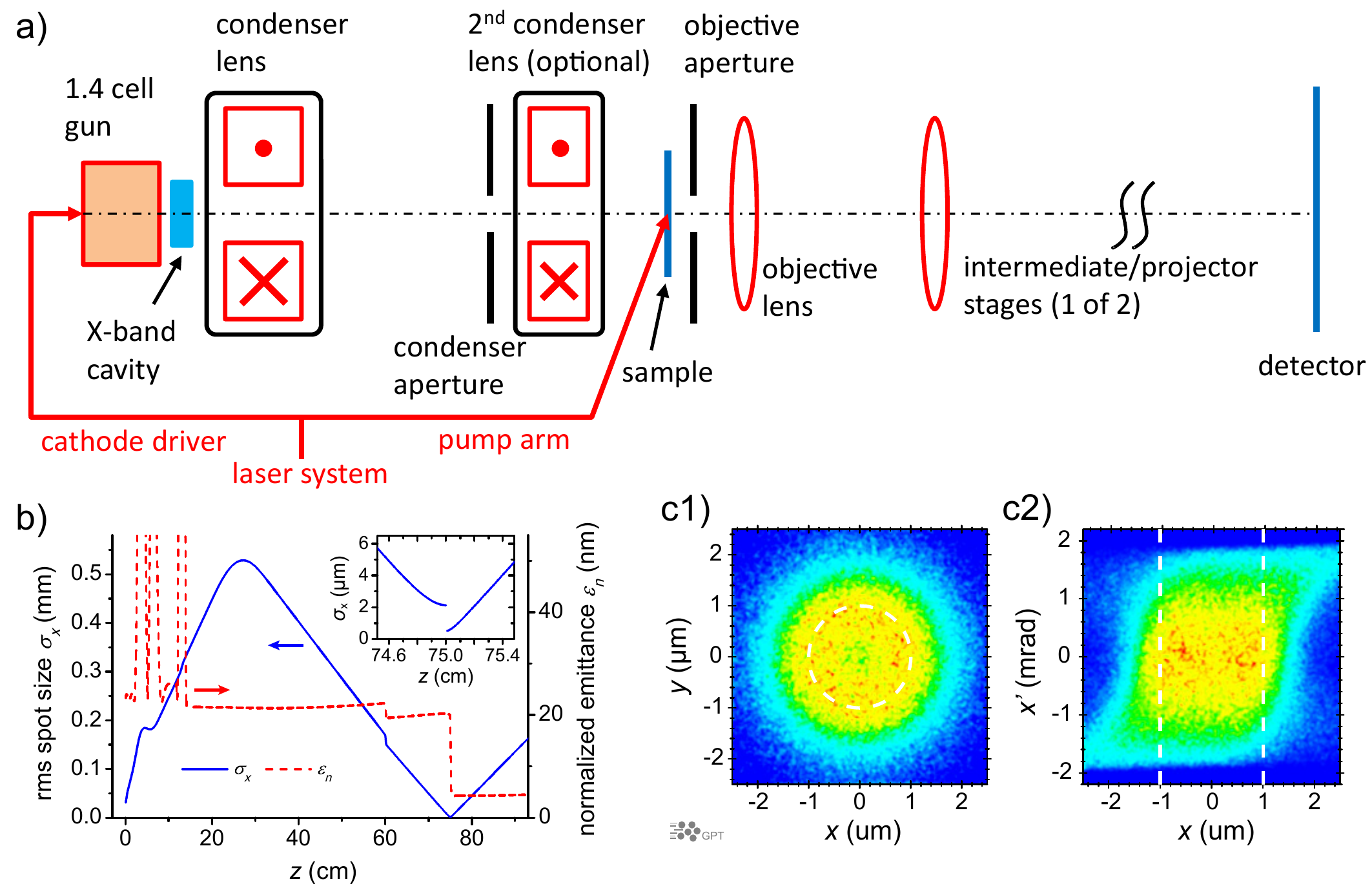}
\caption{(Color online) a) Cartoon concept of the single-shot ps MeV TEM b) The rms spot size $\sigma_x$ and normalized emittance $\epsilon_n$ along the beamline. The inset plot shows $\sigma_x$ evolution close to the sample location at $z=0.75$ m. c1) $x-y$ profile and c2) $x-x'$ distribution of the electron beam at the sample position. Within a 2.0 $\mu$m diameter area, indicated by white dashed lines, the beam has quasi-uniform position and angular distributions.}
\label{fig1}
\end{figure*}

In Fig.~\ref{fig1}(b) we show the evolution of the rms spot size $\sigma_x$ and the normalized emittance $\epsilon_n$ from the cathode at $z=0$ to the sample position at $z=0.75$ m with the condenser lens tuned to deliver a beam waist at the sample position. A collimation aperture is located at 60 cm to block the particles with divergence larger than 2 mrad. The transverse density profile and trace space distribution at the sample plane are shown in Fig.~\ref{fig1}(c1) and (c2). The sample area within a 2 $\mu$m diameter circle is quasi-uniformly illuminated by an average electron flux $200e$/(10 nm)$^2$ and rms beam divergence is $\sigma_\theta = 1.0$ mrad with a quasi-uniform $x'$ distribution.

An optimized set of parameters of the electron source design and simulated beam parameters are reported in Table~\ref{tab:table2}. The improvements discussed in this section, including a shorter (1.4 cell) gun design, cigar aspect-ratio beam shaping, and a higher harmonic ($X$-band) cavity based correlated energy spread compensation, show the feasibility for an RF photoinjector design to satisfy the beam illumination requirements for single-shot ps MeV TEM.

\begin{table}[h]
\caption{\label{tab:table2}%
Parameters for the operation of the electron source and the electron beam qualities at the sample.}
\begin{ruledtabular}
\begin{tabular}{lc}
Parameters & Values \\
\hline
gun gradient & 120 MV/m \\
gun phase & 73 degree \\
initial beam charge & 6.0 pC \\
UV spot size, rms (spherical) & 20 $\mu$m \\
UV pulse duration, FWHM & 10 ps \\
$X$-band cavity gradient & 43.3 MV/m \\
\multicolumn{2}{l}{\bf At the sample (within the 2.0 $\mu$m diameter area)} \\
beam charge & 1.0 pC \\
transverse momentum spread $\gamma\sigma_\theta$, rms & $9.5\times10^{-3}$ \\
bunch length, FWHM & 10 ps \\
kinetic beam energy & 4.4 MeV \\
relative energy spread $\delta\gamma/\gamma$, FW50 & $1.5\times10^{-5}$ \\
\end{tabular}
\end{ruledtabular}
\end{table}

\section{Design of the electron-optical column}

The downstream section of the electron-optical column consists of several lenses that image the beam from the sample (object plane) to the detector (image plane). In most TEMs, this section typically includes 3-6 stages, with an objective lens, one or more intermediate lenses, and one or more projector lenses. The spatial resolution of TEMs is strongly dependent on the optical properties of these lenses, in particular on the spherical aberration $C_s$ and chromatic aberration $C_c$ of the objective lens closest to the sample, where the beam divergence is largest. To minimize $C_s$ and $C_c$ and achieve high spatial resolution, generally very strong objective lenses are utilized~\cite{Haider:Nature1998,Krivanek:Nature1998}. Minimization of the objective focal length $f$ is also important to obtain large magnification ratio maintaining a reasonable total length for the entire microscope.

\subsection{Permanent magnet quadrupole as the lens}

Solenoidal coil is the most commonly used type of electron lens due to the simple axially symmetric geometry. The focal length of a solenoid is $f=[K_s\sin(K_sL)]^{-1}$, where $L$ is the effective magnetic length, $K_s=e_0B_0/(2 \gamma \beta m_0 c_0)$, and $B_0$ is the field strength. For $\gamma=10$ electrons, assuming $B_0=2.2$ Tesla, we have $f=1.6$ cm for $L=2.5$ cm. Such lenses are technically feasible but are very heavy and bulky. For example, the objective lens of the 3 MV electron microscope at Osaka University is $>1$ tons in weight~\cite{osaka3mv}. The field strength of a normal conducting solenoid can hardly be further increased due to the saturation of pole-piece materials. Superconducting solenoids are a viable technology to reach higher field amplitudes as in large bore devices fields as high as 40 Tesla have been demonstrated~\cite{highfield_scmagnets}. Nevertheless these magnets have long effective lengths and significant research and development is required in order to adapt this technology for use in building low-aberration strong lenses for MeV TEMs.

An alternative approach is to use quadrupole magnets. Compared with solenoidal lenses, quadrupoles are much more effective in focusing high energy electrons since the magnetic field is perpendicular to the beam path and the magnetic field component of Lorentz force is maximized. The focal length of a quadrupole magnet can also be expressed as $f=[K_q\sin(K_qL)]^{-1}$. Here $K_q=\sqrt{e_0G/ \beta \gamma m_0 c_0}$, where $G=B_0/a$ is the field gradient, and $B_0$ and $a$ are the tip magnetic field strength and radius of the pole-piece.

Among various kinds of quadrupoles, the permanent magnet quadrupole (PMQ) is a convenient, strong and compact type, widely used for proton or heavy ion beams or for high energy electrons anywhere ultra-strong focusing is demanded~\cite{halbach,pmq1,pmq2,Lim:pmq}. The field gradient can be as high as a few hundred T/m with a typical permanent magnetic material of $B_0=1.2$-$1.4$ T. The effective length is typically set by its physical thickness and the aperture diameter $2a$ which can be as short as a few mm. In what follows we'll consider an electron column design based on PMQs, but most of the considerations can be extended to designs utilizing different kind of quadrupoles.

For example, recent exciting advancement in nanofabrication technique has made possible the development of $\mu$-magnets, i.e. electromagnets with sub-mm dimensions~\cite{harrison:jmems}. For these devices, the pole-pieces are shaped and deposited by nanofabrication techniques with sub-$\mu$m accuracy and the current flows in nano-printed electric stripes. The advantages are that the apertures can be very small hence gradient very high (up to 3000 T/m), and that the focusing strength can be tuned - in contrast to PMQs - simply by adjusting the current in the coils. This technology can also be extended to obtain different magnetic configurations (sextupoles, octupoles etc.) and has the potential for a revolutionary miniaturization of electron optical elements.

Now we discuss our strategy in the design of the electron column. Quadrupole magnets focus the beam in one transverse plane and defocus in the other. At least three PMQs are required to simultaneously fulfill the requirements of imaging in both $x$ and $y$ planes and with equal magnifications. In the linear optics transfer matrix formalism these requirements can be mathematically expressed as $R_{12}=0$, $R_{34}=0$, and $R_{11}=R_{33}$. For the initial optimization we used a hard-edge model with parameters listed in Table~\ref{tab:table3} derived from RADIA magnetostatic simulations~\cite{radia} using permanent magnetic material with residual magnetization $B_0=1.40$ T and an aperture diameter of 2.0 mm. Since the strength of each PMQ is not tunable, the optical properties of the entire triplet are controlled by changing the inter-spacing between PMQs. In Table~\ref{tab:table3} we report the positions of the elements in the triplet that provide equal magnification imaging in $x$ and $y$ planes. The object plane and image plane are located at $z_o=0$ and $z_i=0.2$ m, respectively.
\begin{table}[h]
\caption{\label{tab:table3}%
Parameters of the PMQs for a single triplet imaging stage.}
\begin{ruledtabular}
\begin{tabular}{ccccc}
Name & Thickness & Gradient & Position \\
\hline
$Q_1$ & 6 mm & 506.9 T/m & 6.74702 mm \\
$Q_2$ & 6 mm & -506.9 T/m & 14.92282 mm \\
$Q_3$ & 3 mm & 537.4 T/m & 21.67476 mm \\
\end{tabular}
\end{ruledtabular}
\end{table}

The transfer matrix between $z_o$ and $z_i$ is
\begin{equation}
R =
\begin{pmatrix}
-11.8 & R_{12} & 0 & 0 \\
-57.0 & -8.46\times10^{-2} & 0 & 0 \\
0 & 0 & -11.8 & R_{34}\\
0 & 0 & -73.8 & -8.46\times10^{-2} \\
\end{pmatrix}.
\end{equation}

The $R_{12}$ and $R_{34}$ terms can be infinitely small in ideal numerical solutions. With state-of-the-art piezo-based high precision control of the PMQ positions at a few nm level, these two terms can be made $<1\times10^{-6}$ m/rad. More detailed tolerance studies should include the tilt, rotation, and strength error of the PMQs.

It is possible to calculate the aberrations of the optical system by looking at the higher order terms in the transfer matrix. In particular, we find $C_{c,x}\equiv T_{126}/M=14$ mm, $C_{s,x}\equiv U_{1222}/M=8.6$ mm and $C_{c,y}\equiv T_{346}/M=48$ mm, $C_{s,y}\equiv U_{3444}/M=95$ mm, where $M\equiv-R_{11}=-R_{33}=11.8$ is the magnification. These values can be obtained directly from the output file of a high order transport code such as COSY INFINITY~\cite{cosy}, and have been verified by fitting the results of single-particle GPT tracking simulations.

\begin{figure}[htb]
\includegraphics*[width=85 mm]{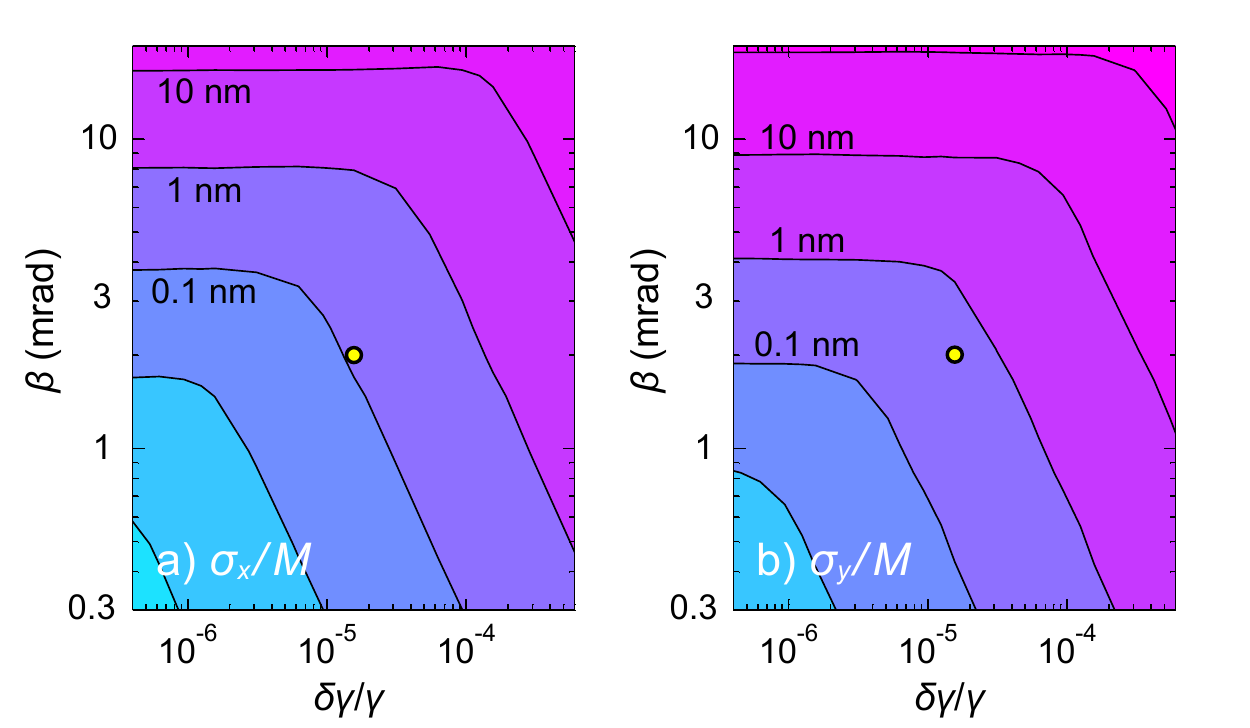}
\caption{(Color online) GPT tracking results of the dependence of the FW50 image size (a) $\sigma_x/M$ and (b) $\sigma_y/M$ of a point source, on its FW50 relative energy spread $\delta\gamma/\gamma$ (Gaussian distribution) and the objective collection semi-angle $\beta$ (uniform angular distribution).}
\label{fighardedge}
\end{figure}

In Fig.~\ref{fighardedge} we show how the relative beam energy spread $\delta\gamma/\gamma$ and the collection semi-angle $\beta$ affect the image size (beam spot size on the image plane) of a point source. The image size value has been converted back to the object plane by dividing the magnification. These results indicate that the blur in the image induced by aberrations can be kept below 1 nm if we limit the collection semi-angle to 2 mrad and the energy spread at low 10$^{-5}$ level, in agreement with the back-of-the-envelope estimates discussed in Section II. For larger values chromatic and spherical aberrations quickly increase the beam size and degrade the imaging performances. The difference in $x$ and $y$ plane are due to the order of the quadrupole orientation in the triplet. By rotating 90 degrees all of the PMQs, we reverse the horizontal and vertical aberration coefficients.

\begin{figure}[htb]
\includegraphics*[width=85 mm]{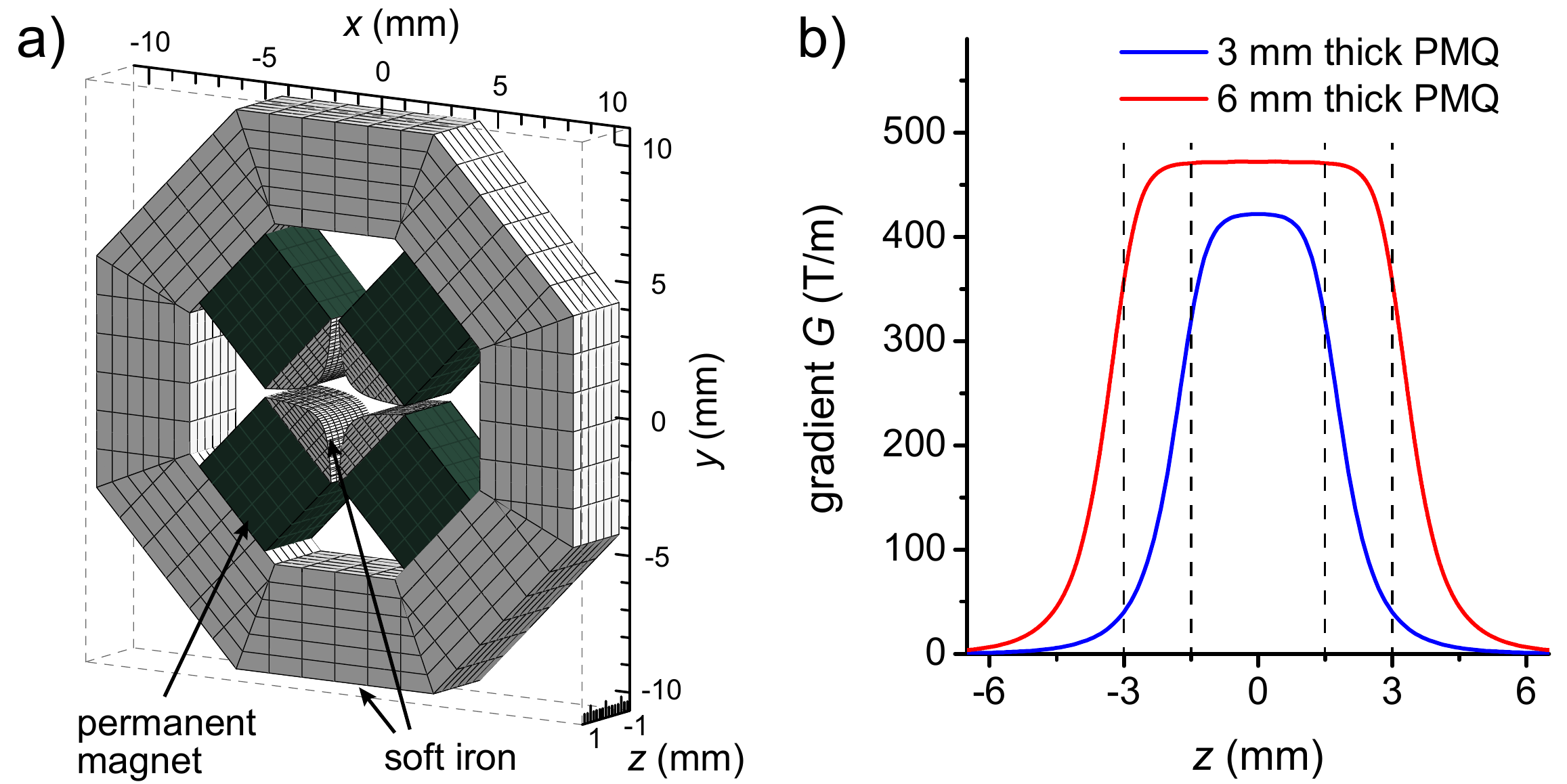}
\caption{(Color online) (a) RADIA model of a 3 mm thick PMQ magnet and (b) the calculated on-axis focusing gradient of a 3 mm and a 6 mm PMQs. Dashed lines indicate the physical boundaries of the 3 mm and 6 mm thick PMQs.}
\label{figpmq}
\end{figure}

Practical PMQs have fringe fields extending beyond their physical boundaries and might have higher order multipole components (for example, octupole component) other than the ideal quadrupole moment. These effects may cause notably different aberrations compared to those predicted based on the ideal hard-edge model. To evaluate these effects, we performed GPT tracking using fully three-dimensional (3D) field maps of PMQs. The 3D PMQ field maps were generated using the RADIA model and imported into GPT. The physical model of the PMQ, including cubic permanent magnet blocks, soft iron yoke and parabolic-shape pole-pieces, is shown in Fig.~\ref{figpmq}, together with the calculated on-axis focusing gradient.

\begin{figure}[htb]
\includegraphics*[width=85 mm]{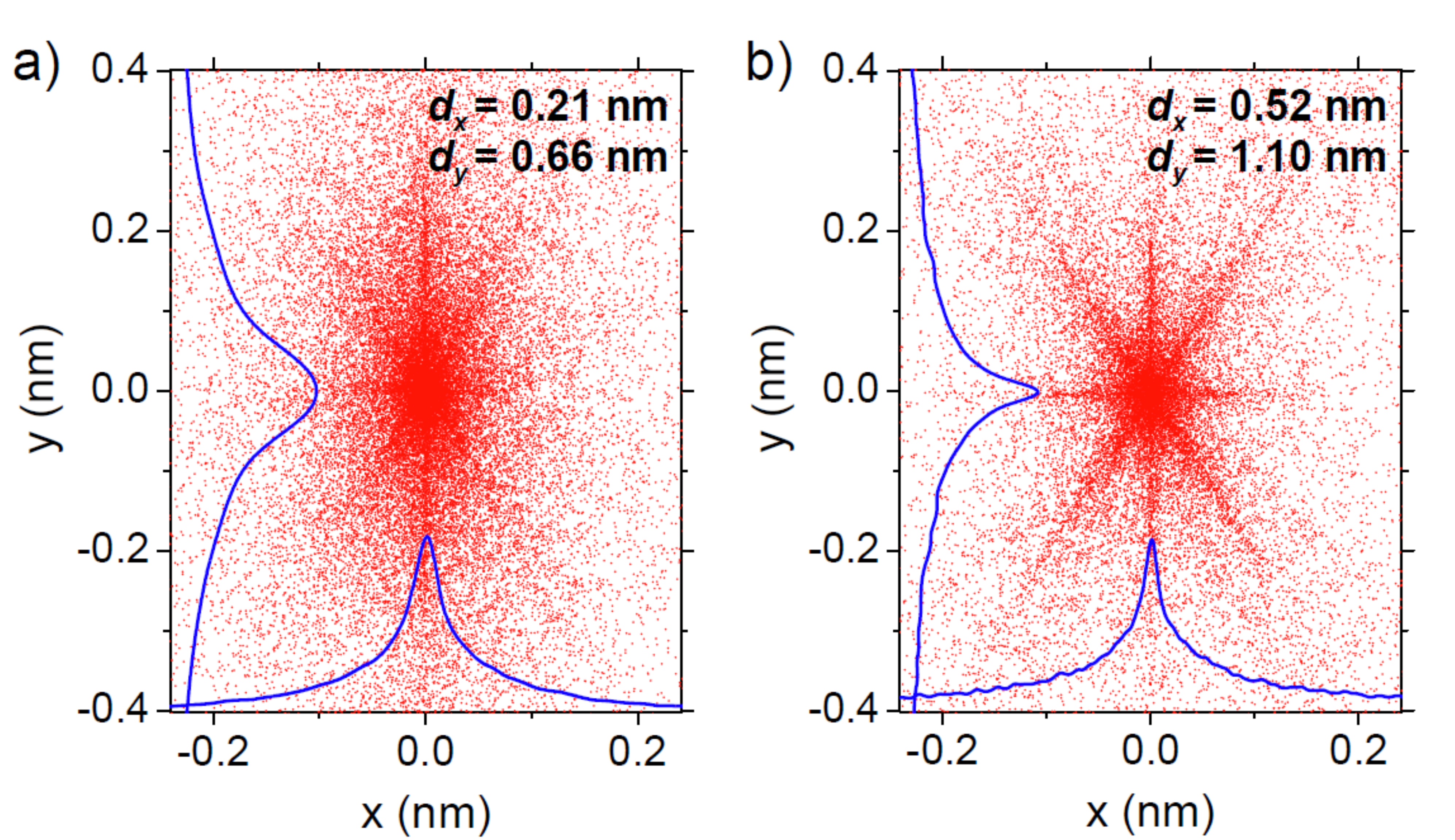}
\caption{(Color online) Image disk profiles of a point source modeled using (a) the hard-edge ideal quadrupole models and (b) the 3D field map from RADIA. The collection semi-angle is 2 mrad and FW50 beam energy spread is $1.5\times10^{-5}$. Transverse dimensions have been referred back to the object plane.}
\label{fig3dmapspot}
\end{figure}

Compared with the hard-edge model, 3D field map results show a reduced magnification, from 11.8 times to 11.5 times, due to the overlap of the fringe fields hence partial cancellation of the focusing strengths of adjacent PMQs which have opposite polarizations. By fitting the GPT particle tracking results we also obtained the aberration coefficients. Chromatic aberrations in both $x$ and $y$ planes are only increased by ~10\% compared with the hard-edge model. However, spherical aberrations are larger by a factor of 5-6 due to the residual octupole component in the field map. In Fig.~\ref{fig3dmapspot}(b) we show the image disk of a point source with the collection angle set equal to the beam divergence listed in Table~\ref{tab:table3}. The eight-fold cross feature related to the octupole field component in 3D field maps is clearly visible. The FW50 disk size referred back to the object plane is still $\lesssim1$ nm since a large fraction of the electrons is still concentrated in the bright central spot.

Dark field imaging using only electrons scattered to larger angles could also be considered. The disk size stays $<$5 nm with the collection semi-angle increased to 3 mrad. Further reduction of the $C_s$ by optimizing the pole-tips shapes, and whether it is possible to decrease the $C_s$ to even below the ideal hard-edge quadrupoles values, i.e. building $C_s$-corrected PMQ triplet, are interesting topics for future studies.

The image size of a point source here is not necessarily the spatial resolution of the microscope. Shot-noise due to finite electron flux and $e$-$e$ interactions within the beam, which will be discussed in the next section, may be larger limiting factors to the spatial resolution. Finally, it is worth noting that the wave-like property of electrons were not included in these calculations. For $\gamma=10$ electrons (de Broglie wavelength 0.25 pm) and a few mrad collection angle, the diffraction limited resolution is on the order of 1 {\AA} thus has negligible effect on the spatial resolution of the instrument.

In our conceptual design, the complete imaging column includes two more stages - an intermediate stage and a projector stage - after the objective triplet lens. The intermediate stage uses the same PMQ triplet as the objective lens and magnifies 30.7 times in 0.5 m. PMQs in the projector stage need to use larger apertures to accommodate the magnified beam spot size hence have reduced focusing gradient. The projector stage magnifies 28.4 times in 1 m. The magnification of the entire column is $11.5\times30.7\times28.4=1.0\times10^4$ times. A 10 nm area and the 2 $\mu$m illuminated region of the sample will be imaged to 100 $\mu$m and 2 cm at the image plane, respectively, comfortably accommodated by the spatial resolution and field of view of state-of-the-art high efficiency detector for MeV electrons~\cite{renkai:jap}.

\subsection{Simulation of the imaging process}

We then simulate the image formation process of a test target through the column under the optical setting we discussed above. The test target consists of four groups of horizontal and vertical line pairs similar to those in the widely used USAF 1951 target. The four groups have line width and spacing of 15 nm, 10 nm, 5 nm, and 2.5 nm, respectively. Bright field imaging mode is considered here. We assumed the electrons hitting the bars are unscattered thus have an transverse momentum spread $\gamma\sigma_\theta=9.5$ mrad, same as the illumination beam listed in Table~\ref{tab:table3}. Electrons hitting other regions of the test target were assumed to be scattered to 5 times the illumination angle. An aperture was set up at 3 mm after the sample, and the aperture size allowed all the unscattered electrons to go through while $\sim$96\% of scattered electrons were blocked. The beam energy spread was assumed to stay at the same level since the energy loss in most samples will be a very small fraction of the MeV kinetic energy.

\begin{figure}[htb]
\includegraphics*[width=85 mm]{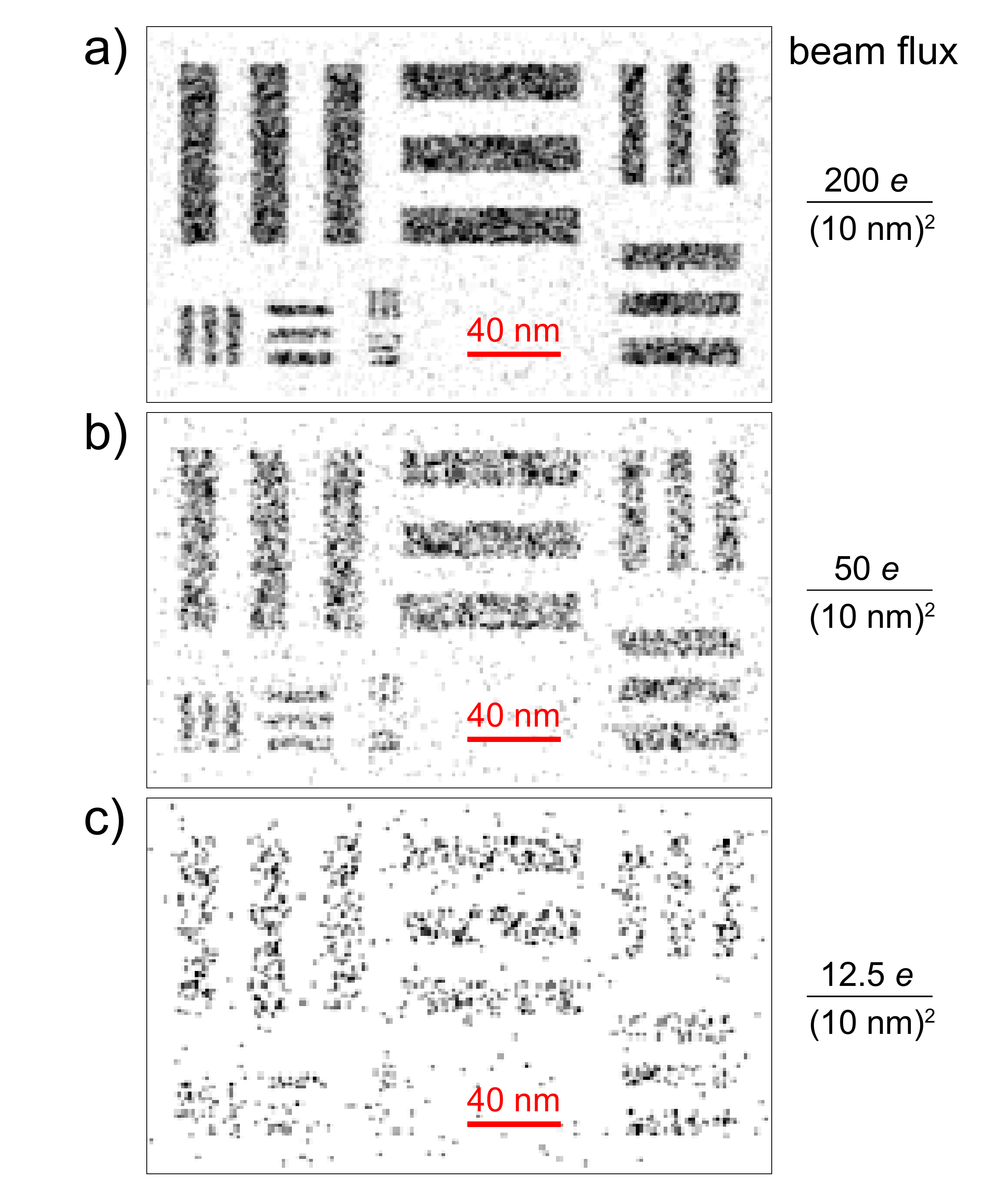}
\caption{(Color online) Simulated electron beam images of the test target under beam illumination flux levels of (a) $200e/(\textrm{10~nm})^2$, (b) $50e/(\textrm{10~nm})^2$, and (c) $12.5e/(\textrm{10~nm})^2$.}
\label{targetnosc}
\end{figure}

In Fig.~\ref{targetnosc} we show the simulated images for three illumination flux levels of (a) $F=200e/(\textrm{10~nm})^2$, (b) $50e/(\textrm{10~nm})^2$, and (c) $F=12.5e/(\textrm{10~nm})^2$. Otherwise identical beam parameters are used for the three images. The relative intensity fluctuation of a feature in the image is $\frac{1}{l\sqrt{F}}$, where $l\times l$ is the feature size and $F$ is the beam flux. As either the beam flux or feature size becomes smaller the increased intensity fluctuation gets close to the contrast in the image, which is close to unity for the test target. The visibility of the features in Fig.~\ref{targetnosc} can be used to estimate the limit on the spatial resolution.

\section{Electron-electron interactions}

In the previous section discussing the design of the electron optical column, $e$-$e$ interactions were not taken into account, hence the size of the image disk of a point source was completely determined by the collection angle, beam energy spread, and the intrinsic aberrations of the electron lenses. In reality, $e$-$e$ interactions may distort or even wash out the information of the sample imprinted in the beam phase space as the beam propagate from the sample to the detector plane. In this section we will evaluate these effects.

$e$-$e$ interactions in a beam can be represented by the sum of the smooth space charge forces and the stochastic scattering resulting from pairwise discrete particle interactions~\cite{reiser}. In order to evaluate the smooth space charge effects, the electron beam can be treated as a non-neutral plasma 'fluid' with continuous charge density distribution. The electric and magnetic fields in this case can be calculated by integration over the entire charge density distribution and are also smooth functions in space and time. The stochastic scattering term is due to collisional events when the motion of a particle is primarily affected by one or few of its nearest neighbours rather than the collective field of the entire beam.

In principle, both components of $e$-$e$ interactions are inherently included in full-scale (i.e. one marco-particle in simulation for one real electron) particle tracking using pair-wise interaction model and could be precisely modeled taking advantage of recent remarkable advances in scientific computing. Nevertheless it still requires a significant amount of computation resources and time to track through the microscope column a beam with ~$10^7$ electrons even for a single run. For example, a single run with only $10^4$ particles requires $\sim$50 hour$\cdot$cores. Furthermore, multiple runs are necessary to reveal the scaling with relevant parameters and guide the design and optimization of the microscope.

\begin{figure}[htb]
\includegraphics*[width=85 mm]{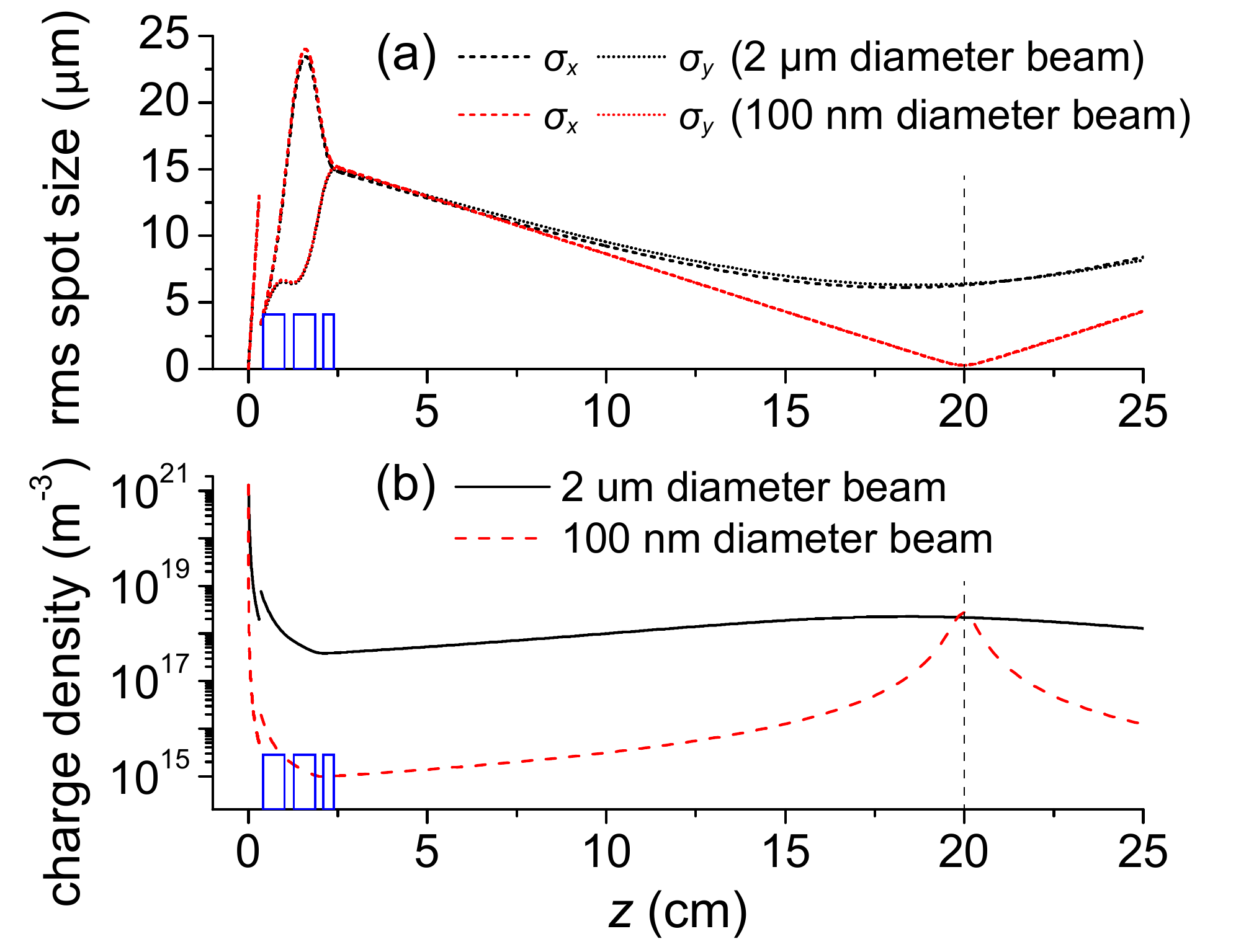}
\caption{(Color online) (a) rms spot size $\sigma_x$ and $\sigma_y$, and (b) charge density evolution in the first stage of the column of the full beam (2 $\mu$m initial diameter, black) and a transversely scaled beam (100 nm initial diameter, red). The blue rectangles represent PMQs positions of the objective lens.}
\label{figdensity}
\end{figure}

Unfortunately, it is also not possible to simply scale\footnote{Here we refer to the scaling in the transverse direction. Longitudinally scaled simulation will be discussed in Section~\ref{sec:stochastic}.} the simulation, that is, only simulating a small transverse portion of the beam which has the same charge density evolution as the full beam, to significantly reduce simulation time. To illustrate this point, we show in Fig.~\ref{figdensity} the beam spot size and charge density in the first stage of the column for the full beam (2 $\mu$m diameter) and a scaled beam with 100 nm diameter spot size. The two beams have same initial current density, energy spread, and divergence.

The mesh-based '\texttt{spacecharge3Dmesh}' algorithm in GPT is suitable to model the transverse profile evolution of the full beam~\cite{reiser}. For the scaled 100 nm beam the '\texttt{spacecharge3Dmesh}' algorithm and a full-scale simulation using the pairwise '\texttt{spacecharge3D}' algorithm (which is now feasible due to the reduced number of particles) yield almost identical (within ~2\%) results. The charge densities of the two beams are exactly the same at the sample plane ($z_o=0$). In a large portion of the column the beam spot size is determined by initial divergence which is same for both beams. For the scaled beam its charge density is much smaller thus the effects of $e$-$e$ interactions are severely under-estimated, roughly by a factor of $g_t^2$ with $g_t$ being the transverse scaling factor $g_t=2~\mu\textrm{m}/100~\textrm{nm}=20$. The charge density for the two cases become roughly equivalent again only close to the image plane ($z_i=20$ cm) where the transverse profiles of both beams are roughly magnified by the same amount. In fact, space charge effects have a notable effect in the full beam case and generate a shift of the image plane and are responsible for the small difference in charge density close to $z_i$.

These considerations led us to the development of alternative new strategies to calculate the effects of $e$-$e$ interactions in the column. We will discuss how the smooth space charge defocusing forces, mostly the nonlinear part, affect the imaging condition in Section~\ref{sec:smooth}. The effects of the stochastic scattering will be considered in Section~\ref{sec:stochastic}.

\subsection{Smooth space charge effects}
\label{sec:smooth}

The strategy to calculate the effects of smooth space charge forces can be summarized in three steps: i) calculate the evolution of the density profile of the full beam in the column. This can be done using a relatively small number of macroparticles ($\sim$$10^4$, compared to $\sim$$10^7$ in a full-scale simulation) using a mesh-based space charge algorithm; ii) calculate the smooth space charge field map within the beam, based on the charge density evolution, at any position in the column; iii) track the motion of each individual electron in the smooth space charge field map, superimposed with the PMQ fields.

\begin{figure}[htb]
\includegraphics*[width=85 mm]{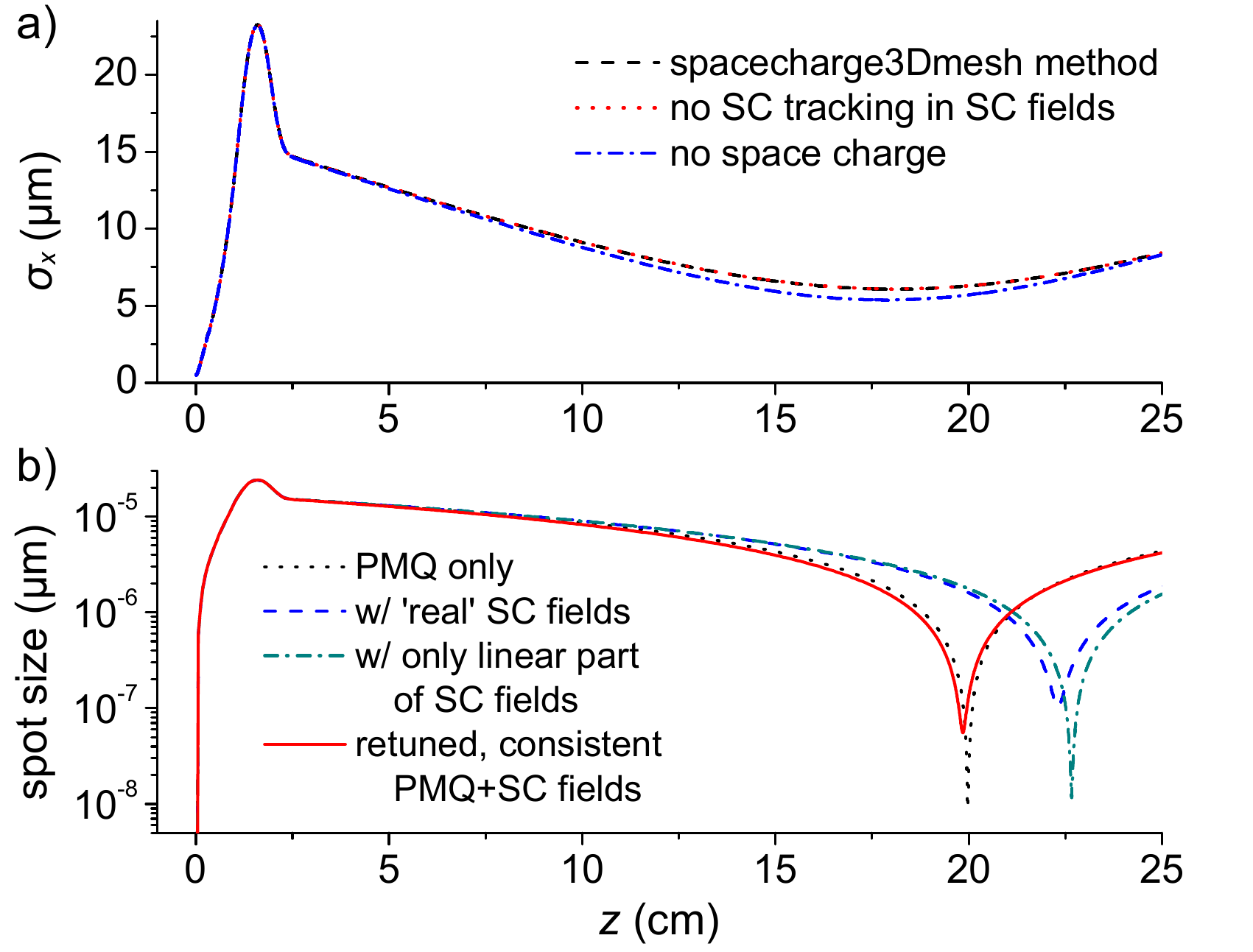}
\caption{(Color online) (a) Comparison of the spot size obtained (black dash) directly using the '\texttt{spacecharge3dmesh}' method and (red dot) the one using superposed PMQ and space charge field map without turning on any $e$-$e$ effects in GPT. The result (blue dash-dot) with only the PMQ field is also shown. (b) Spot size evolution of a point source (black dot) with only PMQ field tuned at imaging at $z_i=20$ cm, (blue dash) after including the smooth space charge field, and (cyan dash-dot) with only the linear component of the smooth space charge field. (Red solid) PMQ triplet is retuned to image at $z_i$ again with new space charge field calculated under the new PMQ setting.}
\label{sccomparison}
\end{figure}

In its rest frame, the beam has an elongated shape (beam aspect-ratio $A=\gamma\sigma_z/\sigma_{x,y}\gg 1$), thus the space charge electric fields are predominately transverse. The magnitudes of the transverse electric fields $E_{x,y}$ scale with the beam current density. For example, for a long, transversely uniform elliptical beam, $E_{x,y}$ at a point $(x,y)$ can be written as
\begin{equation}
E_{x,y}=\frac{I(r_x,r_y)}{\pi\epsilon_0c_0\beta}\cdot\frac{x(y)}{r_{x,y}(r_x+r_y)}\\
\label{uniformellipsefields}
\end{equation}
where $r_{x,y}$ are the semi-axis of the smaller ellipse passing by point $(x,y)$, and $I(r_x,r_y)$ is the enclosed beam current. The ratio $r_x/r_y=\sigma_x/\sigma_y$, where $\sigma_x$ and $\sigma_y$ are the rms spot sizes of the full beam. Equation \ref{uniformellipsefields} is strictly valid only for a uniform density beam but can be used as a first approximation to describe the fields of any elliptical charge distribution~\cite{lblnote}. Electric repulsion forces are partially cancelled by magnetic attraction, thus the total space charge forces acting on the particles can be represented by $E^s_{x,y}=E_{x,y}/\gamma^2$. We fit $I(r_x,r_y)$ using polynomial functions\footnote{The polynomial fitting of $I$ uses even order components (due to middle-plane symmetry) up to 10th order and precisely represents the smooth distribution. For example, when $10^4$ macro-particles are used in simulation, the difference of $I$ between direct counting of macro-particles and polynomial fitting is $<1$\%, on the same magnitude of the macro-particle shot-noise.} (up to 10th order). Smooth space charge field map is then calculated based on the smooth polynomial representation of $I$ using Eq.~\ref{uniformellipsefields} and imported into GPT as external three-dimensional electric field map. Particles are tracked in the superimposed space charge field and PMQ field without turning on any $e$-$e$ interactions in GPT. In Fig.~\ref{sccomparison}(a) we compare the spot size $\sigma_x$ of the full beam simulated using this technique with that obtained directly using the '\texttt{spacecharge3Dmesh}' method. The good agreement between the two curves provide evidence of the validity of this technique.

It is then possible to track particles from a point source at high precision to evaluate the aberrations associated with the smooth space charge field. The black dot line in Fig.~\ref{sccomparison}(b) shows the FW50 spot size of the beam from a point source under only the PMQ field. When smooth space charge field map of the full beam is included, the position of the new 'image plane' has significantly shifted downstream, consistent with our understanding that space charge forces are defocusing. Further, the size of the 'image' is significantly larger due to the strong aberration introduced by the space charge field.

If the transverse beam profile is uniform, i.e. the fitting result of $I(r_x,r_y)$ only contains 2nd order terms, the associated space charge forces will be perfectly linear functions of transverse positions minimizing the aberration. This is confirmed in simulation. When only the linear part of the space charge field is included in particle tracking, the image plane is shifted but the size of the image disk is only slightly increased, as shown by the dark cyan dash-dot curve in Fig.~\ref{sccomparison}(b). After smooth space charge field map is included, the PMQ focusing needs to be adjusted (strengthened) to image again at $z_i$, hence the space charge field map need to be recalculated. A few iterations allow imaging at $z_i$ in a self-consistent way with a new PMQ setting and the space charge field calculated under the same new PMQ setting. The red solid line in Fig.~\ref{sccomparison} shows the spot size under the new optical and space charge field settings.

\begin{figure}[htb]
\includegraphics*[width=85 mm]{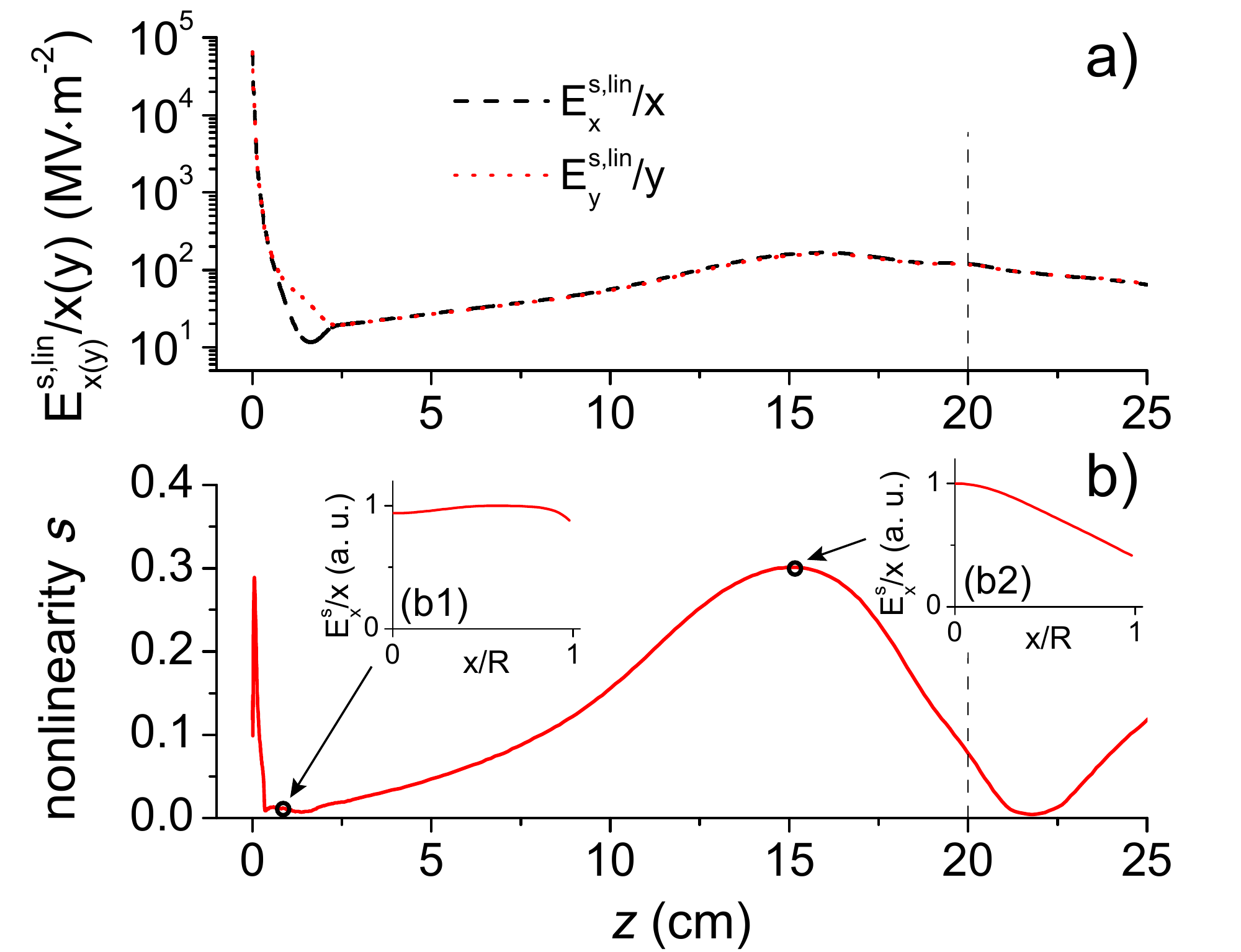}
\caption{(Color online) (a) Field gradients of the linear part of in the space charge field. (b) Nonlinearity $s$ of the space charge field. The field gradient $E^s_x/x$ at two positions where nonlinearity is (b1) low and (b2) high are shown.}
\label{gradandresi}
\end{figure}

Linear space charge defocusing forces do not notably degrade the image disk, but the integrated space charge defocusing effect has to be smaller than PMQ focusing to ensure the formation of cross-over and imaging. The field gradients of the linear part in the space charge field $E^{s,lin}_{x(y)}/x(y)$ are shown in Fig.~\ref{gradandresi}(a). For example, $10^4$ MV/m$^2$ defocusing field gradient needs to be counter-balanced by 33 T/m quadrupole field or 0.26 T solenoid field.

Nonlinear components in space charge fields are directly related to the nonuniform density profile of the beam. The lowest order nonlinear component of the density distribution scales as $x^4$ and $y^4$, and the associated space charge forces have $x^3$ and $y^3$ dependence, contributing to particle motion as dynamic spherical aberrations. Higher order nonlinear components in density and field distributions also exist but at smaller magnitudes.

As discussed in Section III, the beam distribution at the sample is approximately uniform both in density and in angle, so at least initially the space charge field is mostly linear. The non-uniformity in density distribution mainly originates from the evolution of the beam initial state along the column. At each $z$ position in the column the transverse coordinate of a particle can be written (at first order) as the sum of $R_{11}x$, related to its initial position and $R_{12}x'$ proportional to its initial divergence. Whenever one of these terms dominates over the other one, the beam distribution is uniform and the resulting space charge field is linear, but when the two terms are comparable the beam transverse profile is given by the convolution of the initial position and angle distributions and becomes strongly non-uniform with a decreasing density at its edge.

We quantify the nonlinearity using a parameter $s=\int s(x)dx/\int dx$, where $s(x)=|I^{\textrm{full}}(x)-I^{\textrm{2nd}}(x)|/|I^{\textrm{full}}(x)|$, and $I^{\textrm{full}}(x)$ and $I^{\textrm{2nd}}(x)$ are the complete polynomial fitting results of the beam shape and only the 2nd order (uniform) component, respectively. $s$ value in the first stage of the column is shown in Fig.~\ref{gradandresi}(b). The beam density is only roughly uniform close to the object and image planes, and characterized by strong nonlinear components at other positions.

The effects of nonlinear space charge forces can be quantified using equivalent spherical aberration coefficients $C_3^s$ and $C_5^s$, related to the 3rd and 5th orders space charge field, respectively. These coefficients can be extracted by fitting GPT particle tracking results for the position on the image plane of a single particle with a varying initial angle at the object plane. For example, with the current design we have $C_{x,5}^s=-2.5\times10^6$ m, $C_{x,3}^s=2.4$ m, and $C_{y,5}^s=-4.5\times10^6$ m, $C_{y,3}^s=5.0$ m. The magnitudes of these aberration coefficients are determined by the transverse profile (relative importance of nonlinear components) and the average current density of the beam. It might be possible to minimize these coefficients by varying the PMQ focal length, stage length, beam spot size and divergence at the sample, as well as the imaging mode (bright-field or dark-field). An intriguing research opportunity is to explore octupole- and dodecapole-based correction modules to minimize the smooth space charge spherical aberrations in an independent and flexible way.

\begin{figure}[htb]
\includegraphics*[width=85 mm]{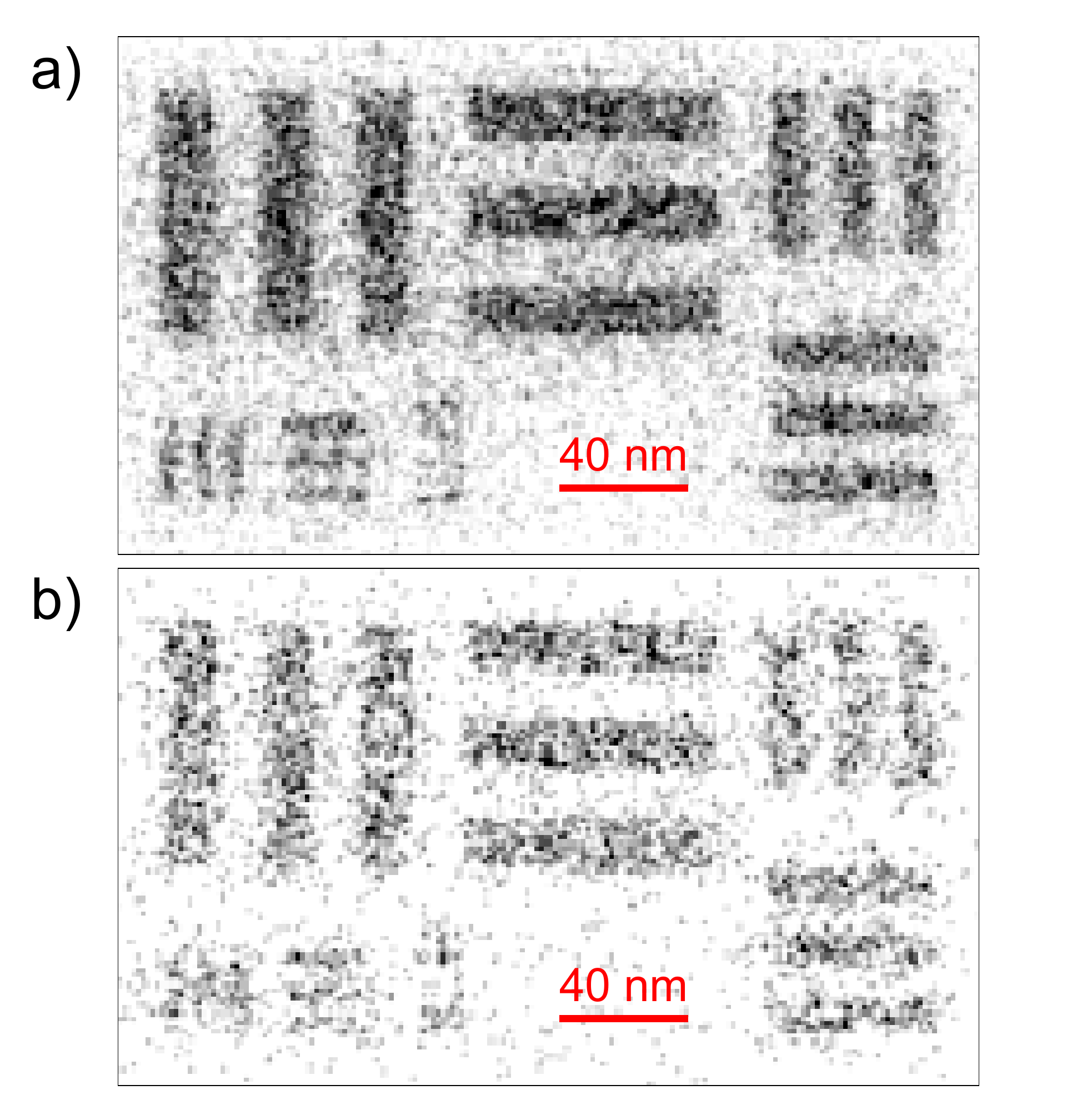}
\caption{(Color online) Simulated image of the test target for (a) $200e/(10~\textrm{nm})^2$ and (b) $50e/(10~\textrm{nm})^2$ illumination flux, with smooth space charge effects taken into account.}
\label{imagecomparison}
\end{figure}

Finally, we applied the above computation scheme to visualize the image of the test target. In Fig. \ref{imagecomparison} we show the images for (a) $200e/(\textrm{10~nm})^2$ and (b) $50e/(\textrm{10~nm})^2$ illumination flux at the end of the first stage with dimensions converted back to the object plane. Space charge aberration coefficients in the second stage are up to ten times larger than those of the first stage. But due to the $M$ times smaller divergence at the entrance to the second stage (hence $M^3$ times smaller image disk size scaling) and $M$ times larger 'object' size, smooth space charge effects have negligible impact on image quality in the second and following stages of the column. As shown in Fig.~\ref{imagecomparison}, 10 nm line pairs can be clearly distinguished in both $x$ and $y$ planes. The visibility of 5 nm wide line pairs are notably degraded compared to the result when smooth space charge effects are not included. We note that the smooth space charge aberration is proportional to beam current density and the shot-noise limited resolution is inversely proportional to the square root of beam flux. Thus for a given bunch length, there exists an optimum condition for the spatial resolution which is given by the beam current for which the contributions from smooth space charge effects and shot-noise are comparable.

\subsection{Stochastic Scattering Effects}
\label{sec:stochastic}

Stochastic scattering between close-by particles can lead to an increase in the beam divergence along the column which results in a degradation of image resolution. First principle pairwise $e$-$e$ interaction algorithms need to be used to precisely model this effect. Longitudinally scaled simulation, i.e. simulating a full-transverse size but a thin longitudinal slice of the beam, can correctly predict the $e$-$e$ interaction effects if the aspect-ratio of the slice stays $\gg1$ and the space charge forces remain predominantly transverse. Scaling down the bunch length by a large factor $g_l=\tau_\textrm{full}/\tau_\textrm{slice}$ (for example up to 100), the computation time becomes manageable even using pairwise algorithm, but the beam flux on the screen will be much smaller than reality by a factor of $g_l$ preventing us from the possibility to visualize the image due to shot-noise considerations. $\tau_\textrm{full}$ and $\tau_\textrm{slice}$ are the length of the full beam and the slice, respectively. To this end, we adopt the 'image-disk convolution' approach to model the image and evaluate spatial resolution under $e$-$e$ interactions as described below.

\begin{figure}[htb]
\includegraphics*[width=85 mm]{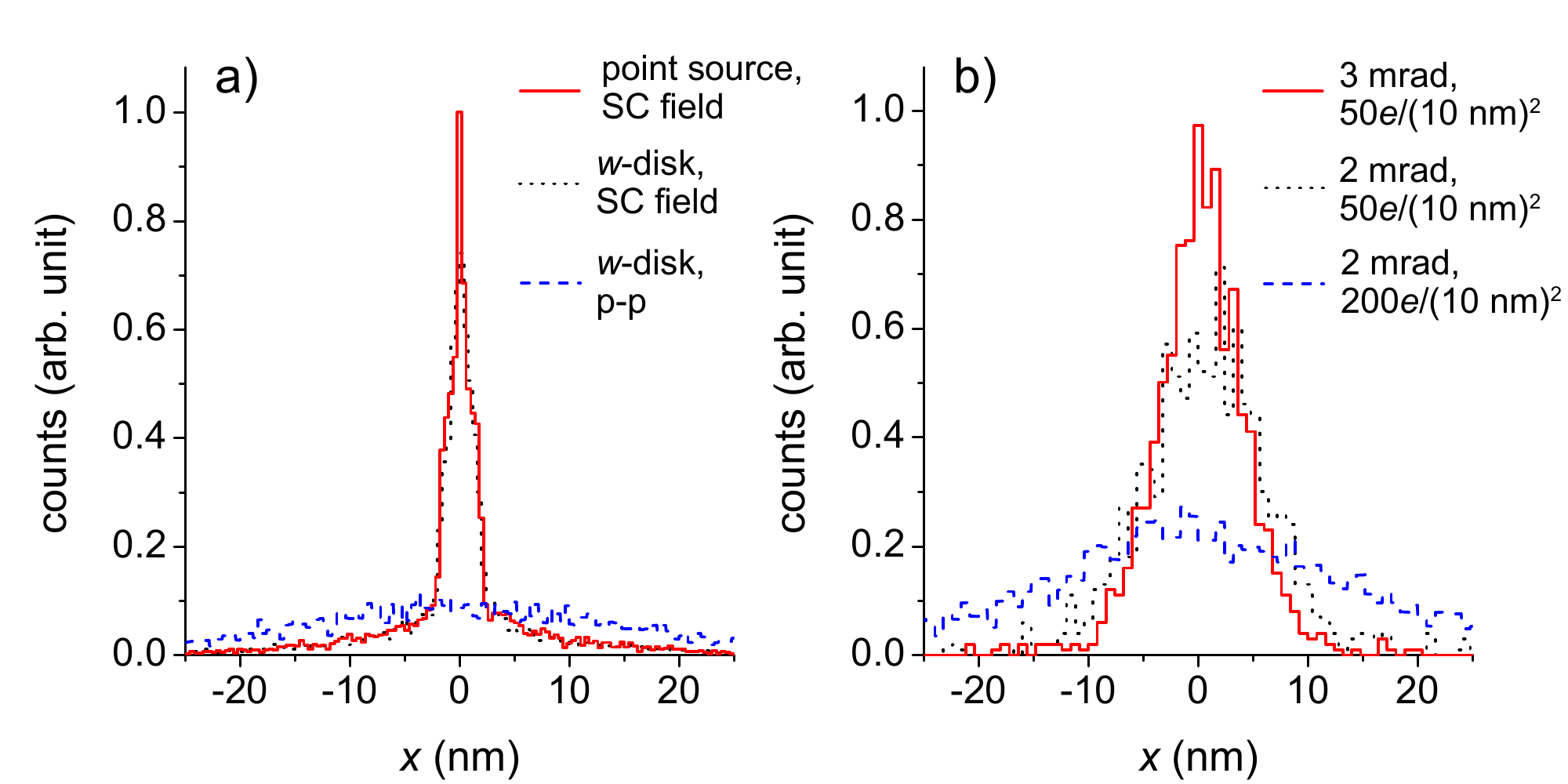}
\caption{(Color online) (a) Histogram of the image of a point source (red solid) and $w$-disk of a full-transverse-size beam (black dot) tracked in the smooth space charge field, together with the $w$-disk of a full-transverse-size beam tracked using first principle pairwise $e$-$e$ interaction model (blue dash). (b) $w$-disk of a full-transverse-size beam tracked using first principle pairwise $e$-$e$ interaction model with divergence and illumination flux of (red solid) 3 mrad, $50e/(10~\textrm{nm})^2$, (black dot) 2 mrad, $50e/(10~\textrm{nm})^2$, and (blue dash) 2 mrad, $200e/(10~\textrm{nm})^2$.}
\label{imagediskhist}
\end{figure}

If there is an ideal imaging from the object plane to the image plane, then $w\equiv-Mx_o-x_i$ will be zero for each particle, where $M$ is the magnification and the minus sign before $M$ is due to the fact that the image is reversed. In reality, aberrations due to both PMQs and $e$-$e$ interactions spread $w$ to a disk of finite size on the image plane. The shape of the $w$-disk should quantitatively agree with the image disk of a point source, as the two distributions essentially both describe how a point is imaged to a finite size due to the presence of aberrations. This is confirmed, as shown in Fig.~\ref{imagediskhist}(a), where we show the profiles of the image of a point source (red solid) and the $w$-disk of a full-transverse-size (2 $\mu$m diameter) beam in good agreement with each other. These two profiles are simulated using the smooth space charge fields discussed in Section~\ref{sec:smooth}. $200e/(10~\textrm{nm})^2$ illumination flux and 2 mrad collection semi-angle were assumed. Roughly 25\% of particles go through the objective aperture located at 3 mm. Finally, by convolving the $w$-disk or image-disk profile with the ideal image of the test target, we can compute the image at the detector plane.

A thin longitudinal slice of the full beam ($g_l = 100$) is simulated in GPT under PMQ fields using the first principle pairwise $e$-$e$ algorithm ('\texttt{spacecharge3D}' method). The simulation in principle includes the effects of both stochastic scattering and smooth space charge. The number of particles used in the simulation is equal to the number of electrons in the slice. Random, instead of Hammersley sequence, distributions in initial positions and angles are used. The $e$-$e$ interactions shift the position of the image plane. The minimal FW50 size of the $w$-disk is actually a good indication to find both the position of the image plane and the magnification value. The result (blue dot) is shown in Fig.~\ref{imagediskhist}(a). Same initial beam conditions as those for the other two curves were used. The result shows that stochastic scattering introduces a large spread of the image disk which notably degrades the spatial resolution.

\begin{figure}[H]
\includegraphics*[width=85 mm]{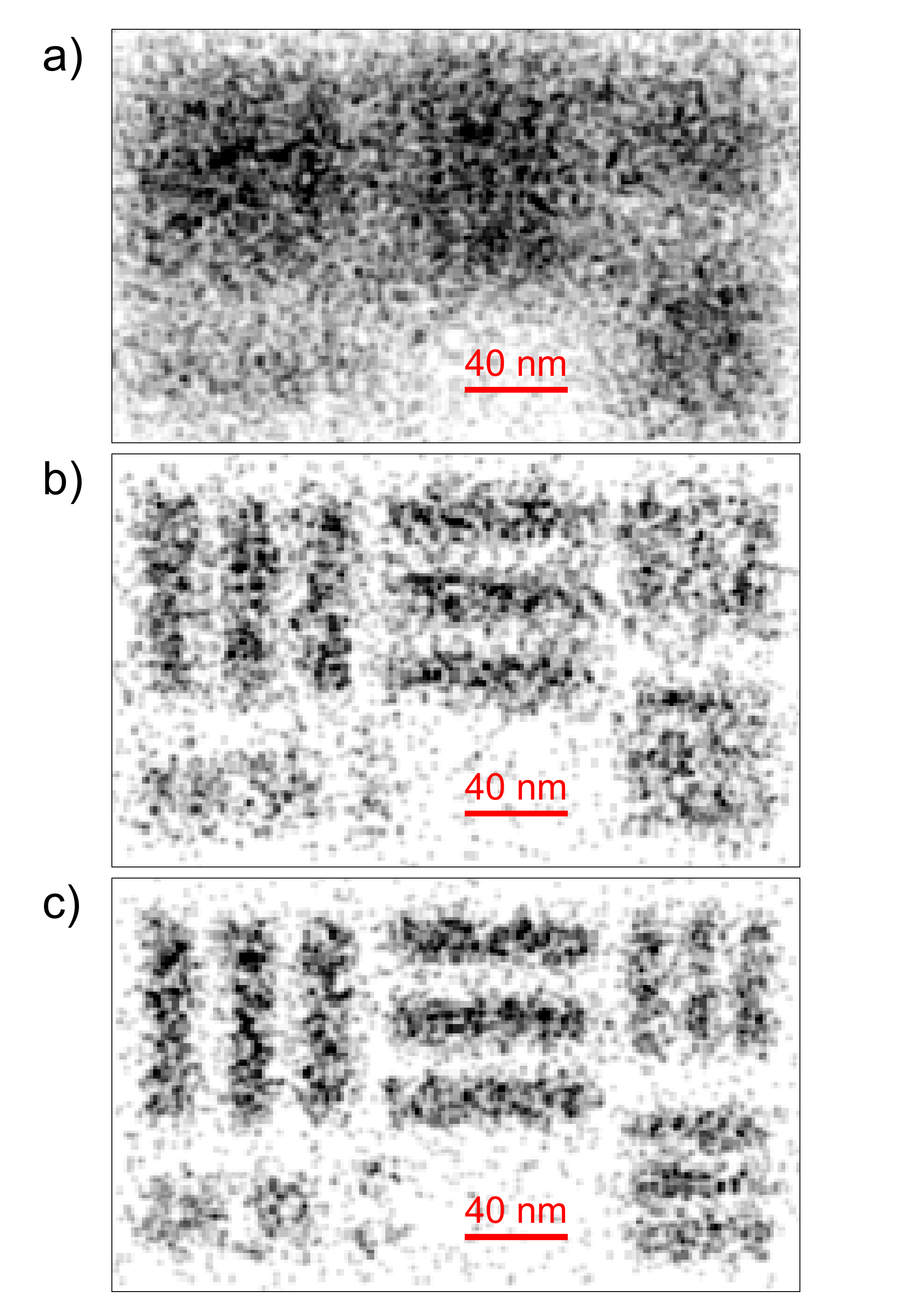}
\caption{(Color online) Images of the test target simulated with first-principle pairwise $e$-$e$ interaction model, for beam divergence and illumination flux of (a) 2 mrad, $200e/(\textrm{10 nm})^2$, (b) 2 mrad, $50e/(\textrm{10 nm})^2$, and (c) 3 mrad, $50e/(\textrm{10 nm})^2$. }
\label{p2pimage}
\end{figure}

A straight-forward approach to reduce the effects of stochastic scattering is to decrease the beam illumination flux, e.g. to $50e/(10~\textrm{nm})^2$ level, which still allows good visibility of 10 nm full width line pairs. As shown in Fig.~\ref{imagediskhist}(b), 4 times reduction in beam flux leads to roughly 2 times decrease in the FW50 size of the $w$-disk. We can further decrease the size of the $w$-disk by increasing beam divergence (assumed matched by collection semi-angle for the bright-field imaging mode), which helps to reduce the charge density after the sample, until geometric aberration of the PMQ triplet prevents the image disk to become smaller. As an example, we show in Fig.~\ref{imagediskhist}(b) that by increasing the beam divergence from 2 mrad to 3 mrad, the FW50 $w$-disk size is reduced from 7.8 nm to 5.4 nm. The convolutions of the the $w$-disks with the ideal image of the test target are displayed in Fig.~\ref{p2pimage}. The images show the possibility to resolve the 10 nm full width line pairs. Using lenses with smaller spherical aberrations which accept larger beam divergence (collection angle) and/or dark-field imaging mode could provide further improvements.

\section{Summary}

In conclusion, we have analyzed various components of a novel MeV-energy time-resolved TEM driven by an RF photoinjector. Innovations in the source design include a higher accelerating field at photoemission, and taking advantage of the cigar regime of operation of RF photoguns. A high frequency RF cavity is proposed to compensate the beam energy spread and minimize the effects of the chromatic aberration. A quadrupole based imaging system is discussed and analyzed with the help of particle tracking simulations. The effects of $e$-$e$ interactions, including smooth space charge forces and stochastic scattering, on the spatial resolution is studied in detail with novel strategies. The final system shows the feasibility of taking single-shot images of samples with 10 ps temporal resolution and 10 nm spatial resolution. This instrument can be useful in the study of materials under extreme conditions, such as the response of materials to laser-induced intense pressure and temperature stimuli. Imaging the motion of dislocations under extreme pressures, which is currently only possible at X-ray FEL in diffraction mode \cite{Milathianaki:Science}, is one of the possible application of this device.

This work was partially supported by DOE Grant No. DEFG02-07ER46272.

\bibliographystyle{unsrt}

\pagebreak

\end{document}